\documentclass[manuscript]{acmart}
\AtBeginDocument{%
  }

\copyrightyear{2025}
\acmYear{2025}
\setcopyright{rightsretained}
\acmConference[LAK 2025]{LAK25: The 15th International Learning Analytics and Knowledge Conference}{March 03--07, 2025}{Dublin, Ireland}
\acmBooktitle{LAK25: The 15th International Learning Analytics and Knowledge Conference (LAK 2025), March 03--07, 2025, Dublin, Ireland}
\acmPrice{}
\acmDOI{10.1145/3706468.3706567}
\acmISBN{979-8-4007-0701-8/25/03}
\begin{document}

\title[Towards Fair and Privacy-Aware Transfer Learning for Educational Predictive Modeling]{Towards Fair and Privacy-Aware Transfer Learning for Educational Predictive Modeling: A Case Study on Retention Prediction in Community Colleges}

\author{Chengyuan Yao}
\email{cy2706@tc.columbia.edu}
\affiliation{%
  \institution{Teachers College, Columbia University}
  \country{USA}
}

\author{Carmen Cortez}
\email{cc4144@tc.columbia.edu}
\affiliation{%
  \institution{Teachers College, Columbia University}
  \country{USA}
}

\author{Renzhe Yu}
\email{renzheyu@tc.columbia.edu}
\affiliation{%
  \institution{Teachers College, Columbia University}
  \country{USA}
}

\begin{abstract}
Predictive analytics is a widely used application of learning analytics, but many resource-constrained institutions lack the capacity to develop their own predictive models or rely on proprietary models trained in different contexts with little transparency. In this context, transfer learning holds promise for expanding reliable and equitable access to predictive analytics, but this potential remains underexplored given existing legal and technical constraints. In this paper, we examine transfer learning strategies in the context of retention prediction at two-year community colleges in the United States, which enroll the most postsecondary students from underserved communities with higher dropout rates than selective universities. We envision a scenario where community colleges can collaborate with each other and with four-year universities to develop retention prediction models under privacy constraints, and evaluate the risks and potential improvement strategies of cross-institutional model transfer for different stakeholders. Using detailed administrative records from 4 research universities and 23 community colleges, which cover more than 800,000 students across 7 cohorts, we first identify performance and fairness degradation when source (external) models are deployed at a target institution without any localization. Fortunately, publicly available institution-level contextual information can be used to forecast these performance drops and offer early guidance for model portability. For model developers under data privacy regulations, sequential training that selects training institutions based on demographic similarities proves useful for enhancing the general fairness of resulting models without compromising performance. For target institutions without local data to fine-tune source models, we find that customizing evaluation thresholds for different sensitive groups is more successful than established transfer learning techniques at improving performance and fairness of deployed models. Our findings suggest the value of transfer learning for more accessible educational predictive modeling and call for judicious use of contextual information in model training, selection, and deployment to achieve reliable and equitable model transfer\footnote{Codes to reproduce our experiments are available at \href{https://github.com/AEQUITAS-Lab/Transfer-Learning-LAK-2025}{https://github.com/AEQUITAS-Lab/Transfer-Learning-LAK-2025}}.

\end{abstract}


\maketitle

\section{Introduction}
With the advancement of digital infrastructure and the abundance of big data in education, predictive analytics has been playing an increasingly important role in various educational decision-making scenarios \cite{namoun2021predicting}. The development of reliable predictive models typically relies on access to high-quality training datasets and careful modeling decisions, but in reality, under-resourced educational agencies with the greatest need for predictive analytics often lack the necessary local infrastructure and resources to build customized models. This dilemma challenges the promise of predictive analytics to alleviate the scarcity of quality educational resources for disadvantaged learners and improve educational equity accordingly.

In this context, transfer learning -- a machine learning technique where a model trained on one task is adapted to perform a different but related task by leveraging existing knowledge \cite{pan2010survey}-- becomes a potential solution to bridging gaps in building predictive analytics between differently resourced schools, institutions, and districts. In fact, many educational organizations have already deployed externally developed predictive models, representing a basic scenario of transfer learning. However, the performance and limitations of such model transfer processes have not been fully understood in educational contexts. Therefore, we present this study that formally investigates transfer learning strategies in one of the most studied educational predictive modeling tasks -- college retention prediction. We focus on fairness and privacy of the technical processes, addressing major ethical concerns for educational stakeholders.

We focus on college retention prediction because college dropout is a persistent challenge in higher education. In the United States, only 39.2\% of students at two-year public institutions graduate within three years, and 65.7\% of students at four-year public institutions graduate within six years~\cite{shapiro2019completing}. As part of their efforts to improve student retention and success, many institutions have utilized retention predictive models to identify at-risk students and inform timely interventions \cite{ameri2016survival, beaulac2019predicting, berens2019early}. As with other predictive analytic tasks in education, under-resourced institutions, such as community colleges, often lack access to high-quality data and technical expertise for in-house model development and in the best case rely on predictive models developed by third-party vendors with limited transparency or local customization~\cite{scull2022community}. In addition, existing privacy and data protection laws, such as the Family Educational Rights and Privacy Act (FERPA), impose restrictions on sharing student data beyond institutional boundaries and add further conditions to the feasibility of transfer learning strategies.

Our study provides a holistic evaluation of ethical issues associated with existing practices of cross-institutional transfer of retention prediction models, as well as potential improvement strategies in line with current regulations. We focus on retention prediction for community colleges given their critical role in serving disadvantaged populations and improving economic and social mobility. Using comprehensive administrative data from 27 institutions across seven student cohorts, we consider risks and transfer learning strategies for both model developers (e.g., third-party vendors) and model users (i.e., community colleges). Formally, we address the following research questions:

\textbf{RQ1.} What are the risks associated with directly applying a pre-trained model for college retention prediction? Under what conditions are these risks higher or lower?

\textbf{RQ2.} For model developers, what strategies during the model training and selection phases may alleviate the risks in model transfer without access to model users’ local historical data?

\textbf{RQ3.} For model users, what strategies during the model evaluation phase may alleviate the risks in model transfer without access to model training details? 

We expect this work to contribute to existing research and practice in a few important ways. First, it will advance the understanding of predictive model portability in education by quantifying the risks associated with cross-context model deployment and identifying strategies to mitigate these risks. Second, it will contribute empirical research on responsible AI by addressing fairness and privacy concerns simultaneously in transfer learning. Third, it will advance more equitable scholarship on college retention prediction by focusing on community colleges which have a higher demand for predictive analytics but are understudied in existing research.

\section{Related Work}
\subsection{Trustworthy Transfer Learning}
Transfer learning has emerged as a powerful technique in the field of machine learning, which enables models to leverage knowledge from one domain or task to improve performance on another. This approach has been particularly successful in fields such as image processing and natural language processing
\cite{jiang-zhai-2007-instance,Fawaz_2018,zhang2019deep}. The core idea behind transfer learning is to utilize the knowledge gained from a well-studied problem with abundant data to address a related problem with limited data. This method not only enhances model performance but also reduces the need for extensive data collection and labeling, which can be both time-consuming and costly \cite{laurer2024less}.

As transfer learning applications expand into sensitive and high-stakes fields such as healthcare, education, and finance, researchers have started to focus their attention on the trustworthiness of the technical strategies in real-world contexts. Two prominent aspects of trustworthiness are fairness and privacy. Algorithmic fairness refers to the principle that machine learning models should provide equitable outcomes and benefits across different demographic groups \cite{mehrabi2021survey}. In transfer learning, the challenge of fairness arises from potential demographic disparities between the source domain, where the model is trained, and the target domain, where it is applied. These disparities can lead to models that perform inequitably for certain groups in the target domain, resulting in biased decision-making \cite{schrouff2023diagnosingfailuresfairnesstransfer}. Various methods have been proposed to mitigate fairness issues in the transfer learning process across different fields \cite{mehrabi2021survey, coston2019fair, mandal2020fairness, rezaei2021robust}, but these approaches largely focus on single sensitive groups, such as race or gender, without addressing the complexities of intersectionality which considers the compound and unique challenges faced by individuals with intersecting identities \cite{kong2022are}.

Privacy poses another ethical and technical challenge within the context of transfer learning. Different fields have distinct privacy constraints. For example, in education, privacy and data protection are regulated by legislation such as the Family Educational Rights and Privacy Act (FERPA), which generally prohibits the disclosure of personally identifiable information from student records without written consent \cite{uscongress1974ferpa}. In general, the primary privacy concern in transfer learning is to preserve the privacy of source domain data. Recent work on source-free domain adaptation (SFDA) addresses privacy and data sharing concerns by adapting a pre-trained model to a new target domain without requiring access to source domain data \cite{yu2023comprehensive}. Researchers have demonstrated its success in computer vision \cite{liang2020shot, wang2021tent}, but the application of this strategy in real-world problems remains limited. In addition, online learning, which involves continuously updating the model as new data become available and allows real-time adaptation to changing data distributions \cite{hoi2018onlinelearningcomprehensivesurvey}, has been used to formulate transfer learning under the term of ``online transfer learning'' \cite{ZHAO201476}. In the process of developing transfer models, this approach can simulate scenarios of cross-context collaboration without violating privacy, such as sharing model weights instead of source data to collaboratively build a more effective transfer model.

\subsection{Transfer Learning in Educational Predictive Modeling}
Model portability has been one of the key challenges in learning analytics \cite{gavsevic2016learning}. Early empirical work revealed how educational predictive models may fail to perform cross different instructional, institutional, and social contexts. For example, \cite{ocumpaugh2014population} found that models for detecting students’ affective states in tutoring systems struggled to generalize across student populations, particularly when rural students were the target population. \cite{li2021limitsalgorithmicpredictionglobe} examined the application of academic achievement prediction models trained on US datasets in less developed countries and found significant performance drops. 

More recently, researchers have started to examine the mechanisms and potential solutions to the issue of model portability, some of whom have formally taken the perspectives of transfer learning. One of the primary barriers to effective model transfer lies in the differences between the source and target contexts, and \cite{xu2024contexts} proposed an analytical pipeline to understand how contextual differences moderate the portability of course-level performance prediction models. In terms of strategies to improve model portability, \cite{swamy2022} explored three transfer learning methods with different model inputs and training paradigms in the context of early success prediction in MOOCs, and found that models with course information had the most satisfactory performance. In our context of college retention prediction across institutional contexts, \cite{luzan2024instance} addressed contextual differences by using an instance weighting strategy to adjust the source model based on the distribution characteristics of the target institution. While this approach improved the overall goodness-of-fit of the transferred model, it requires access to both source and target data, which poses challenges under common privacy constraints. \cite{Gardner_2023} explored a handful of model ensemble strategies across institutions without data sharing and showed some promise in improving model transfer when institutional characteristics are not substantially different.

Despite the existing efforts, the potential of transfer learning is still underexplored in the context of educational predictive modeling. Informed by the accomplishments and limitations of prior research, this study presents a comprehensive evaluation of both ethical risks and improvement strategies when predictive models are transferred across educational contexts.

\subsection{College Retention Prediction}
Over the past few decades, predicting college retention has become a critical area of research in higher education. Researchers have leveraged various data sources, such as institutional administrative data, digital footprints, and survey data, along with cutting-edge machine learning techniques, to identify students at risk of dropping out \cite{yu2021should,bird2021bringing,Aulck_2019,glandorf2024}. Current research on this topic across different countries and institutions has shown promising results in accurately predicting college retention for different student populations at different time points. However, most research is conducted in the context of research-intensive institutions which tend to be more selective and suffer less from retention problems. By contrast, less selective and under-resourced institutions, such as community colleges in the United States, are extremely underrepresented in the literature. This is particularly concerning as community colleges and their counterparts in other countries primarily serve socioeconomically disadvantaged populations such as racial minorities and low-income students who face greater challenges in achieving academic success and are more likely to drop out of college. Accordingly, these colleges are in greater need of the support that retention prediction models can provide. The limited literature in community college contexts has examined performance and fairness in college retention prediction and found that underserved populations may benefit less from predictive models, further underscoring the need for equitable and effective predictive approaches \cite{bird2021bringing,bird2024algorithms}. Related to such challenges, \cite{Gardner_2023} explored cross-institutional transfer of predictive models to inform collaboration between research-intensive universities and under-resourced institutions in model development, although they did not empirically evaluate the approaches in under-resourced community colleges.

\section{Study Context}
We focus on the problem of student retention prediction at community colleges, which may benefit the most from predictive analytics due to their high proportion of minority and underprepared students and low retention rates~\cite{morest2013access}. Similar to~\cite{Gardner_2023}, we envision a collaborative model development paradigm where a community college can deploy externally trained predictive models. A \textit{source institution} is one whose data is part of the training data of the deployed model(s), whereas a \textit{target institution} is the one that deploys the trained model(s). Given our focus on community colleges, target institutions are always community colleges but source institutions can be of any type. There is never data sharing across institutions in line with existing data laws and regulations, although third-party model developers (e.g., a vendor or inter-university consortium) may have simultaneous but separate access to data from multiple institutions in model development. All the analyses in this paper are performed in this context.

\section{Data and Methods}
\subsection{Data Sources}
\subsubsection{Administrative Records}
Through existing research partnerships, we gained access to detailed administrative records at four public research universities located across three states of the United States and 23 community colleges within a southeastern state. These records came from student information systems (SIS) and included individual students' background characteristics as well as academic records in college. For this paper, we restricted the sample to first-time, first-year students entering college in Fall terms between 2013 and 2019, covering over 800,000 students in total. For this study, we purposefully excluded more recent student cohorts whose experience was complicated by the COVID-19 pandemic. Based on these raw administrative records, we created a shared schema of common variables across the 27 institutions\footnote{The shared schema can be found at \href{https://github.com/AEQUITAS-Lab/Transfer-Learning-LAK-2025/blob/main/Table1-Data-Schema.csv}{https://github.com/AEQUITAS-Lab/Transfer-Learning-LAK-2025/blob/main/Table1-Data-Schema.csv}}. 

\subsubsection{Contextual Factors}
We collected institutional-level contextual factors from the Integrated Postsecondary Education Data System (IPEDS) \footnote{\href{https://nces.ed.gov/ipeds}{https://nces.ed.gov/ipeds}}, a comprehensive data collection and analysis system for higher education in the United States, managed by the National Center for Education Statistics (NCES). IPEDS serves as the primary source of detailed information on colleges and universities, covering a wide range of aspects such as enrollment, graduation rates, student demographics, faculty composition, finances, institutional costs, and financial aid. For this study, we included 64 variables across six categories for all 27 institutions: school characteristics, academic compositions, demographic composition, completion rates, cost, and financial aid\footnote{The details of these variables can be found at \href{https://github.com/AEQUITAS-Lab/Transfer-Learning-LAK-2025/blob/main/Table2-Institutional-Contextual-Factors.csv}{https://github.com/AEQUITAS-Lab/Transfer-Learning-LAK-2025/blob/main/Table2-Institutional-Contextual-Factors.csv}}.

\subsection{Prediction Task}
Our central technical task is predicting first-year retention, defined as whether a student who enters an institution for the first time in the fall will re-enroll at the same institution the following fall. This definition aligns with the National Student Clearinghouse’s standard for retention~\cite{gardner2022persistence}. The predictors include student-level variables specified in the shared schema.

In general, we include students entering college between 2013 and 2018 in the training set and reserve those entering in Fall 2019 for testing purposes. Given our focus on community colleges, the 4 research universities are only used for training but not for testing. The specific training and testing data vary across transfer learning strategies explored in this study, as detailed in the next subsection.

\subsection{Transfer Learning Strategies}
We assume that transfer learning has the potential to help community colleges more responsibly take advantage of predictive analytics, as many of them lack the capacity to develop local models or simply rely on vendors' models trained elsewhere. Therefore, we evaluate the following technical transfer learning strategies as well as some of their combinations that can be adopted by target institutions or model developers. As a baseline comparison, 
we also include an ``ideal local'' strategy.

\textbf{Ideal Local}: A local model is trained using historical data from the same target institution on which it is evaluated.

\textbf{Direct Transfer}: A model is trained using data from a source institution and then evaluated on a different target institution.

\textbf{Sequential Training}: The process begins with training a model on the source institution's data. The trained model's weights are then transferred to another institution to perform another round of training. Finally, the model is evaluated on the target institution. To prevent catastrophic forgetting during the sequential training process, we implement the Elastic Weight Consolidation (EWC) strategy \cite{Kirkpatrick_2017}. For instance, EWC helps the model retain prior knowledge (e.g., from the first training institution) while training on subsequent datasets (e.g., the second training institution).

\textbf{Source-Free Domain Adaptation (SFDA)}: SFDA is advantageous for addressing data distribution shifts under privacy constraints, as it enables adaptation to new target domains without directly accessing source data, thereby preserving data privacy. In this study, we explore three benchmark methods within the SFDA framework:

\begin{itemize}
    \item \textbf{Source Hypothesis Transfer (SHOT) \cite{liang2020shot}}: SHOT adapts the feature extractor to the target domain while keeping the classifier fixed, using self-supervised pseudo-labeling and information maximization to align target domain with the source hypothesis.
    
    \item \textbf{TENT \cite{wang2021tent}}:  TENT adapts a model during test time by minimizing the entropy, updating the batch normalization parameters without requiring labeled data.
    
    \item \textbf{Pseudo-Labeling \cite{lee2013pseudo}}: Pseudo-labeling assigns labels to unlabeled target data based on the predicted probability, using these confident predictions to adapt the model to the target domain.
\end{itemize}

\textbf{Customized evaluation thresholds}:
In the model evaluation stage, we propose two customized evaluation threshold methods: one based on the institution’s overall historical outcome statistics (overall-optimal) and the other utilizing demographic-specific thresholds derived from the historical
outcome statistics of individual groups (group-optimal).

In this study, we employed a fully connected neural network model, consisting of a feature extractor, a bottleneck layer, and a classifier, to perform the binary prediction task, which is suitable for all strategies mentioned above.

The workflow of utilizing these strategies is as follows: In RQ1, we evaluate the risks associated with direct transfer by analyzing the contextual similarity (discussed in section 4.4.3) between source and target institutions. For RQ2, we use contextual similarity to guide model selection and employ sequential training to improve model adaptation across institutions. In RQ3, we apply SFDA methods and implement customized evaluation threshold methods to assess and enhance the performance of transfer models in the target institution.

\subsection{Key Metrics}
\subsubsection{Performance and Fairness}
In binary predictive analysis, various performance metrics have been proposed in the literature, each focusing on different dimensions of model evaluation. Widely used metrics, such as those derived from the confusion matrix, require the selection of a specific decision threshold. However, determining a universally applicable threshold for various training and test institutions poses a big challenge, particularly when there are shifts in data distributions. In such cases, the use of a default threshold may lead to suboptimal performance. Therefore, we adopt the \textbf{Area Under the Receiver Operating Characteristic Curve (AUC)} as a primary metric for model evaluation, which is formally defined as: 
\[
AUC(f(\theta)) = \int_{0}^{1} \text{TPR}(\text{FPR}(f_t(\theta))) \, dt
\]
where $t$ is a prediction threshold of the model with the decision rule $f_t(\theta, x) = 1$ if $f(\theta, x) \geq t$, and TPR, FPR stands for true positive rate and false positive rate respectively. AUC scores range from 0 to 1, with higher scores indicating a better ability to distinguish between positive and negative classes, while a random guess will achieve an AUC score of 0.5. AUC offers a threshold-independent assessment of the model's discriminative power by evaluating its performance across all possible thresholds.

To measure the difference in predictive performance under cross-institutional transfer, we employ $\Delta \textbf{AUC}$ \cite{Gardner_2023}, defined as $\Delta AUC(T, T') = AUC(T) - AUC(T')$, where $AUC(T)$ refers to the AUC of a model trained using a transfer scheme $T$. $\Delta AUC$ allows us to quantify the performance difference between various transfer learning schemes. $\Delta AUC = 0$ indicates that the model performs equally well under both schemes. A positive $\Delta AUC$ suggests the model performs better in context $T'$ than in $T$, while a negative $\Delta AUC$ indicates the opposite. Mostly in this study, we evaluate the direct transfer model performance drop compared to the ideally local model, which is $\Delta AUC(\text{local}, \cdot)$.

To evaluate fairness across subgroups and account for intersectionality, we employ the concept of \textbf{AUC Gap}, as introduced by \cite{Gardner_2023}: 
\[
\max_{g, g' \in \mathcal{G}} \left| \mathbb{E}_{\mathcal{D}_k} \left[ f(\theta(\mathcal{D}_{k,g})) \right] - \mathbb{E}_{\mathcal{D}_k} \left[ f(\theta(\mathcal{D}_{k,g'})) \right] \right|
\]
where \(\mathcal{D}_{k,g}\) and \(\mathcal{D}_{k,g'}\) indicate the subset of the data in group \(g\) and \(g'\), respectively. The AUC Gap quantifies the maximum disparity in AUC scores among subgroups, representing the worst-case performance differential within a set of subgroups. In this study, we employ this method to examine the intersectional group based on gender and underrepresented minority (URM) status.

During model evaluation stage, we propose strategies for determining customized thresholds for individual test sets. Since AUC is threshold-independent, we measure the overall model performance with the \textbf{Matthews Correlation Coefficient (MCC)} which accounts for the balance between true positives, false positives, true negatives, and false negatives. In addition, given that retention interventions such as early warning systems often emphasize early identification of students at risk of dropping out, we consider \textbf{Specificity} in our evaluation. Specificity measures the proportion of actual negatives (students who are at risk of dropping out) that are correctly classified by the model.

To evaluate fairness when AUC Gap is not applicable, we use \textbf{Equalized Odds (EO)}, which measures the parity of the true positive rate (TPR) and true negative rate (TNR) between two groups, $A$ = 0 and $A$ = 1. To derive an overall EO score, we compute the EO TPR and EO TNR separately for each group, then take their average to obtain a comprehensive measure of fairness. \[
EO_{(A)} = \frac{| \text{TPR}_{(A=0)} - \text{TPR}_{(A=1)} | + | \text{TNR}_{(A=0)} - \text{TNR}_{(A=1)} |}{2}
\]

In this part, we focus on two demographic categories—race and gender—which are commonly used in evaluating algorithmic fairness. Specifically, we assess fairness between underrepresented racial minorities (URM) vs. non-URM and female vs. male. 

\subsubsection{Distributional Difference}
In this study, we utilize the performance distribution to assess the overall performance across all target institutions. To evaluate the differences between distributions, we employed the \textbf{Wasserstein distance}, which is a measure of the distance between two distributions and is defined as: 
\[
W(P, Q) = \int_{-\infty}^{\infty} \left| F_P(x) - F_Q(x) \right| \, dx
\]
where $F_P(x)$ and $F_Q(x)$ are the cumulative distribution functions (CDFs) of the probability distributions $P$ and $Q$. However, the Wasserstein distance does not have an absolute reference value for evaluation. Existing studies suggest that it should be interpreted relative to the range of the data in order to provide meaningful insights \cite{villani2008optimal}. In this study, we define a 5\% of the data range as the threshold for a "small" Wasserstein distance, known as \textbf{Wasserstein Threshold for Notable Distribution Difference (WTNDD)}. If the WTNDD passed, it indicates no notable difference between the distributions. Conversely, if WTNDD failed, it suggests a notable difference between the distributions.

\subsubsection{Contextual Similarity}
As noted in Section 4.1.2, we collect contextual factors for each institution across six categories. We calculate the similarity scores for each category for each pair of institutions using Gower’s distance \cite{Gower1971}, a metric well-suited for mixed data types. The similarity scores are normalized to fall within the range of 0 to 1 for consistency across categories.

To construct an overall similarity score between source and target institutions, we utilize the results of a regression analysis on AUC Drop (\(\Delta AUC(\text{local}, \cdot)\)) (see Section 5.1 for more details). Specifically, we apply the \textbf{“Coefficient-Significance Weighted Similarity Method”}, which assigns weights to each similarity category based on the regression coefficients and their statistical significance. The weight for each similarity measure \(i\) is calculated as:
\[
W_i = \frac{|\beta_i|}{(1 + p_i)}
\]
where \(\beta_i\) is the regression coefficient for category \(i\), and \(p_i\) is its p-value. This weighting method reduces the influence of variables with lower statistical significance while emphasizing variables with greater impact. 
The weights were normalized so that their sum equals 1. Using these normalized weights, the overall similarity score between source and target institutions is computed as:
\[
X_{\text{overall}} = \sum_{i=1}^{n} W_i \cdot X_i
\]
where \(X_i\) represents the normalized similarity score for category \(i\), and \(W_i\) is the corresponding weight.

\section{Results}
\subsection{Risks of Direct Transfer}

In RQ1, we examine the risks of performance and fairness associated with directly transferring a model developed in a different context. 
\begin{figure*}[h] %
    \centering
    \includegraphics[width=0.9\textwidth]{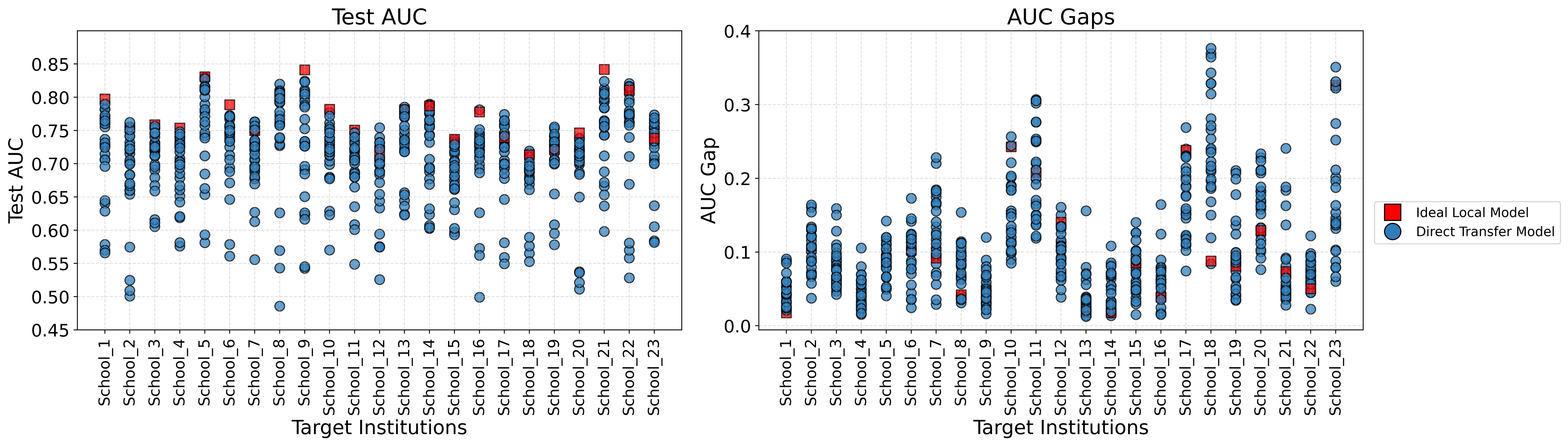} %
    \caption{\textbf{Test AUC (left) and AUC Gaps (right) of pre-trained models for each target institutions. Blue circles represent the tested metrics of pre-trained direct transfer models for each target institution, while red squares indicate the tested metrics of ideal local models.}}
    \label{fig:RQ1.1}
\end{figure*}
Figure~\ref{fig:RQ1.1} (left) shows the AUC for direct transfer models trained on each source institution and applied to each target institution.  The results demonstrate substantial variability in direct transfer AUC performance depending on the source institution. Figure~\ref{fig:RQ1.1} (right) shows the AUC Gap of these models across target institutions. The results indicate that source models face differing levels of fairness challenges.

To assess under which conditions model transferability is more vulnerable, we examine the relationship between contextual similarity and model portability by regressing two metrics - AUC Drop, $\Delta AUC(\text{local}, \cdot)$, and fairness disparity, AUC Gap - on similarity scores across different contextual categories between the source and target institutions. The regression equations are defined as follows:
\[
\Delta AUC(\text{local}, \cdot)= \beta_0 + \sum_{i=1}^{6} \beta_i X_i + \epsilon, \quad \text{AUC Gap} = \beta_0 + \sum_{i=1}^{6} \beta_i X_i + \epsilon
\]
where $X_i$ represents the similarity score of each category as discussed in section 4.4.3. 
\begin{figure*}[h] %
    \centering
    \includegraphics[width=0.7\textwidth]{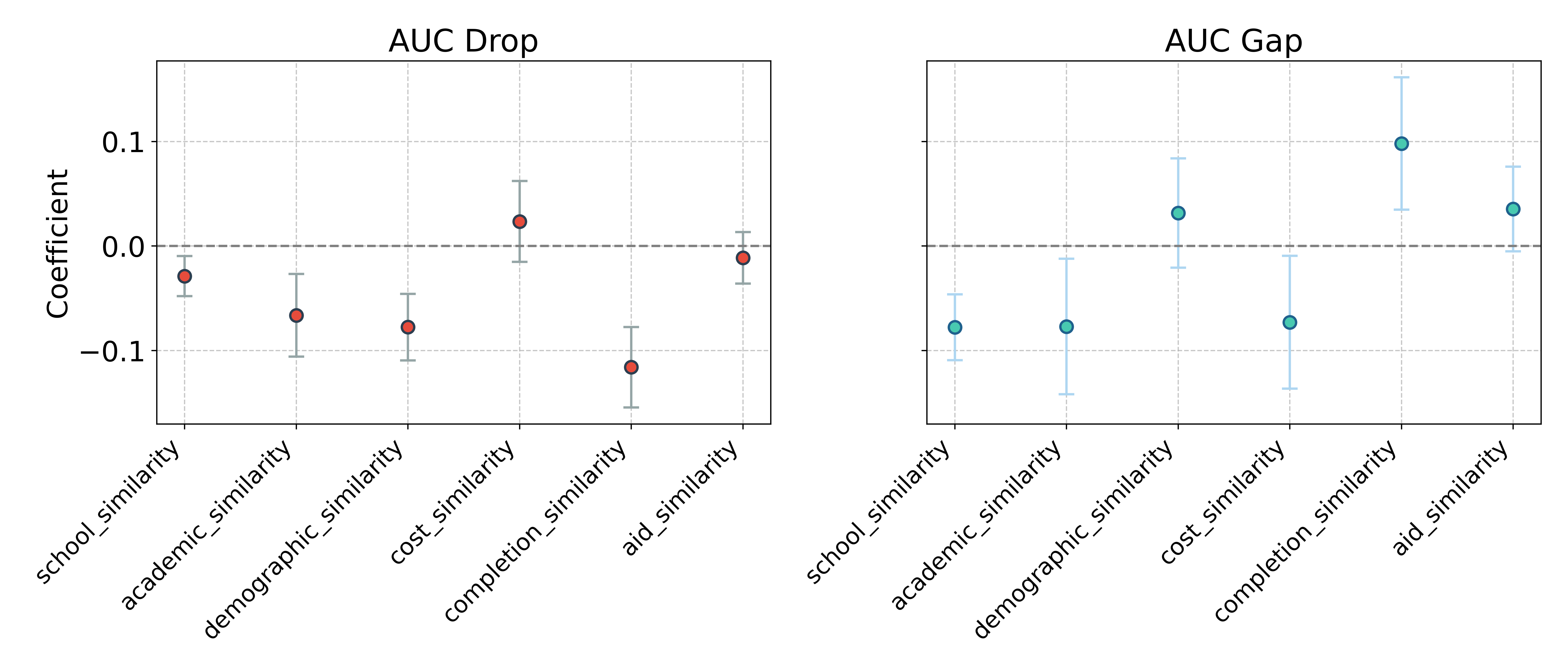}
    \caption{\textbf{Association between contextual similarity metrics and AUC Drop (left) and AUC Gap (right). A negative coefficient indicates a reduction in the AUC Drop or AUC Gap. Error bars represent 95\% confidence intervals.}}
    \label{fig:RQ1.3.1}
\end{figure*}
Figure~\ref{fig:RQ1.3.1} (left) shows that models perform better, with smaller AUC drops, when the source and target institutions are more similar in terms of school, academic, demographic, and completion factors. Figure~\ref{fig:RQ1.3.1} (right) shows that greater similarity between source and target institutions in school, academic, and cost factors reduces the AUC gap, indicating improved fairness. However, higher similarity in completion factors tends to increase the AUC gap. Notably, the regression model for AUC Drop has an R-squared value of 0.600, while the R-squared value for the AUC Gap regression is only 0.056. This implies that while performance degradation is strongly influenced by differences in contextual factors between institutions, fairness change is more complex and less predictable based on the examined variables.

Following the regression results for AUC Drop, we calculate an overall similarity score for each pair of source and target institutions using Coefficient-Significance Weighted Similarity Method discussed in section 4.4.3. Since cost similarity tends to enlarge the AUC drop, we drop it from calculating the overall similarity. 
\begin{figure*}[h] %
    \centering
    \includegraphics[width=\textwidth]{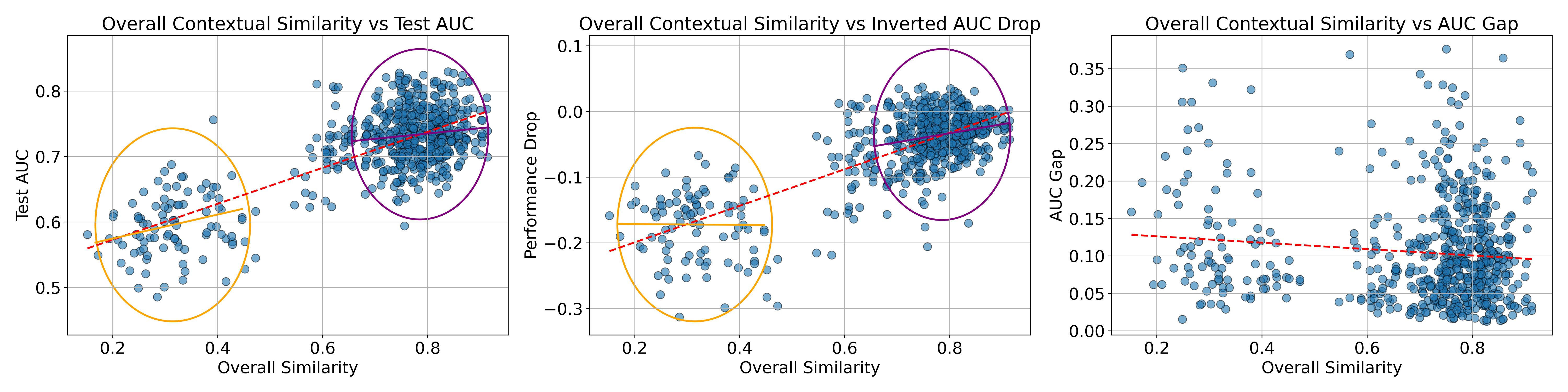}
    \caption{\textbf{Relationship between overall contextual similarity and the model's transferability and fairness. Blue circles represent the tested metric of pre-trained direct transfer models on each target institution.}}
    \label{fig:RQ1.3.2}
\end{figure*}Figure~\ref{fig:RQ1.3.2} illustrates the relationship between overall similarity and three key metrics: Test AUC, Inverted AUC drop, and AUC Gap. The left two plots demonstrate a positive linear trend, indicating that higher overall similarity between institutions is associated with improved AUC scores and a reduced performance drop. Notably, when overall similarity falls below 0.5, the model tends to experience more pronounced performance issues. Additionally, two distinct clusters emerge within the ranges of 0.15–0.45 and 0.65–0.95, each characterized by relatively low internal variance. In contrast, the relationship between overall similarity and fairness, as measured by AUC Gap, exhibits a largely flat trend with a slight negative slope. This suggests that overall similarity exerts minimal influence on AUC Gap, consistent with the coefficient of determination results observed in previous analyses.

\subsection{Transfer Learning Strategies in Model Training and Selection}
From a model developer's perspective, privacy constraints pose challenges in selecting the most suitable pre-trained model that they can pass to the target institution. Based on the previous results, we propose that selecting the most similar school, based on overall contextual similarity, can help identify an appropriate model. The model selected using this method is referred as Most Similar Training Institution(MSTI) model.
\begin{figure*}[h] %
    \centering
    \includegraphics[width=0.7\textwidth]{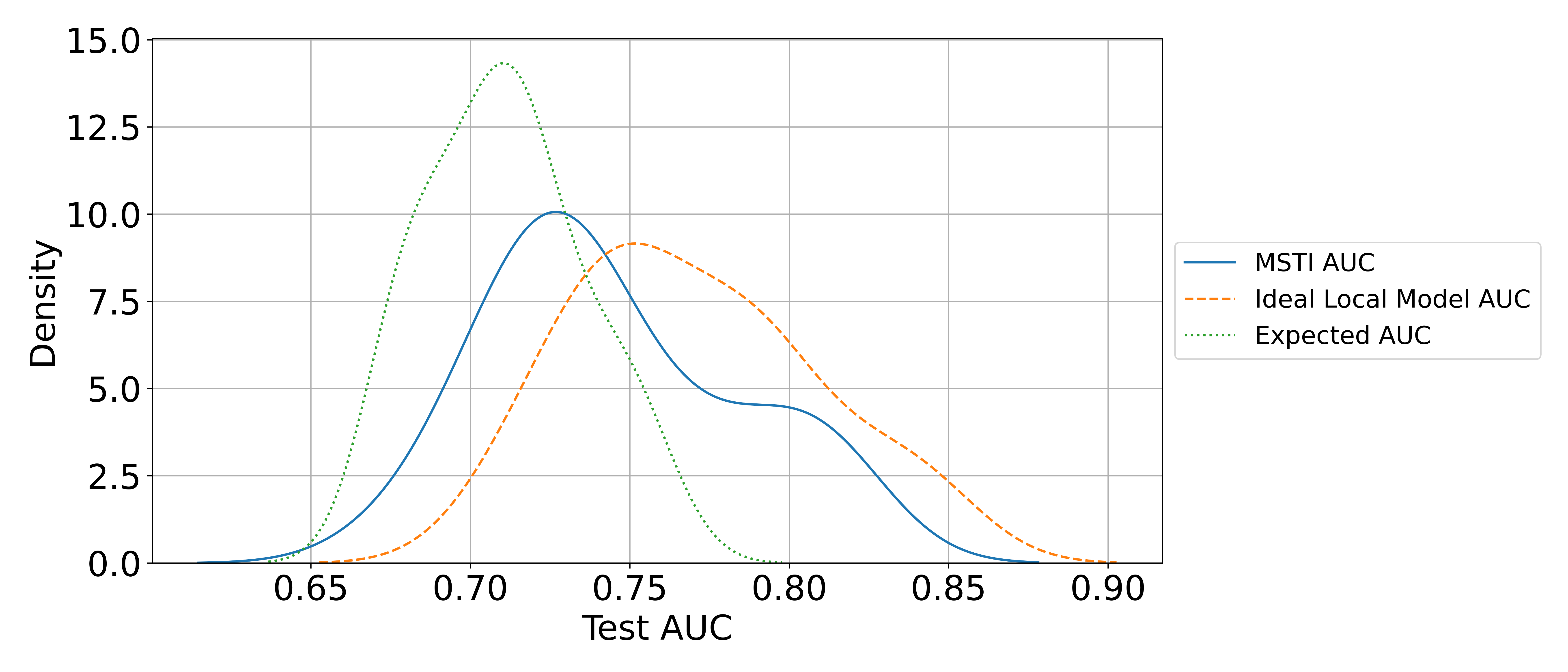}
    \caption{\textbf{Performance distributions across all target institutions in the dataset, comparing Test AUC for three scenarios: MSTI model, the ideal local model, and the expected outcome. The histograms are normalized to have an area of 1 (density).}}
    \label{fig:RQ2.1.2}
\end{figure*}
Figure~\ref{fig:RQ2.1.2} demonstrates that, although the distribution of Test AUC derived from the MSTI models generally underperforms relative to that of the ideal local models (WTNDD failed), it exhibits a marked improvement over the distribution of expected AUC obtained from random model selection (WTNDD failed). This indicates that selecting a model based on contextual similarity can enhance overall predictive performance compared to a randomly chosen model.
\begin{figure*}[h] %
    \centering
    \includegraphics[width=0.8\textwidth]{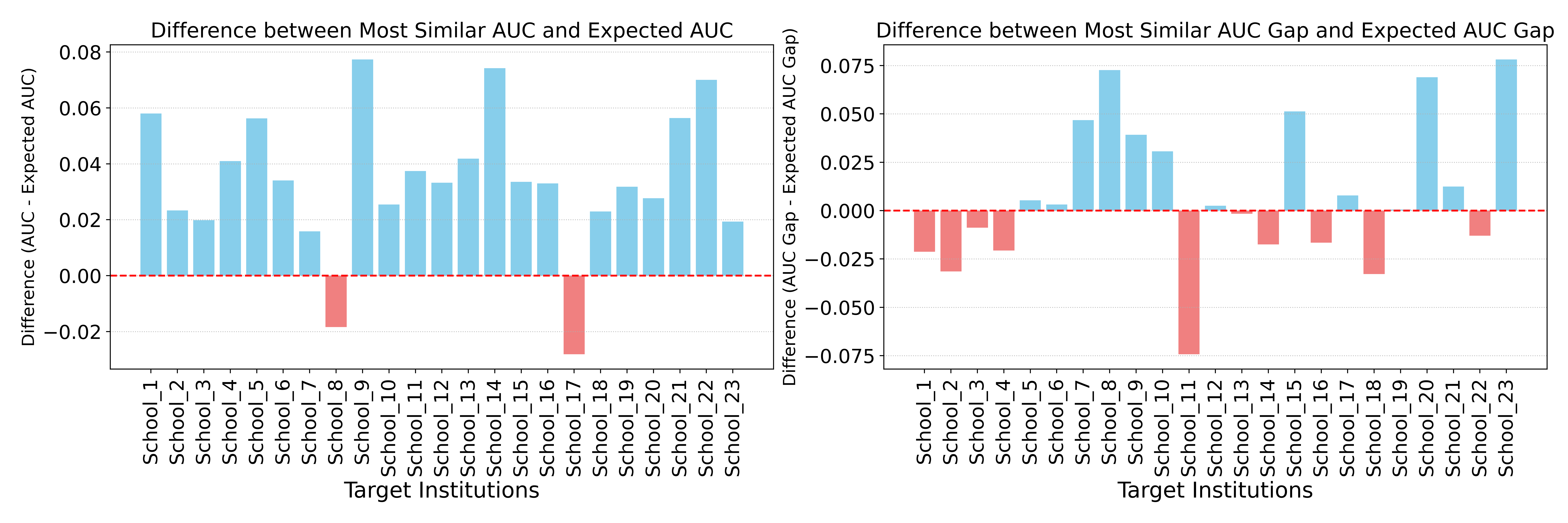}
    \caption{\textbf{Differences between MSTI's Test AUC and AUC Gap with expected values across target institutions.}}
    \label{fig:RQ2.1.1}
\end{figure*} From the perspective of each target institution, Figure~\ref{fig:RQ2.1.1} indicates that the Test AUC from the MSTI model for 21 out of 23 target institutions exceeds the expected AUC. However, no significant improvements in fairness were observed, consistent with our previous conclusions that fairness is less predictable from contextual similarity.

Existing studies have demonstrated that a diverse training set can help address fairness issues \cite{kizilcec2021algorithmicfairnesseducation}.  However, privacy constraints prevent model developers from combining data from multiple institutions to create a more diverse training dataset. To tackle this challenge, we propose a potential solution aimed at improving model fairness while respecting privacy constraints.
First, for each target institution, we firstly identify the MSTI model. Next, among other possible training institutions with an overall similarity greater than 0.6—chosen to avoid performance drops—we use demographic similarity to select the most demographically dissimilar institution, labeled as Training Institution 2. Finally, we conduct sequential training on MSTI and Training Institution 2. The obtained model is referred as sequential model. 
\begin{figure*}[h] %
    \centering
    \includegraphics[width=\textwidth]{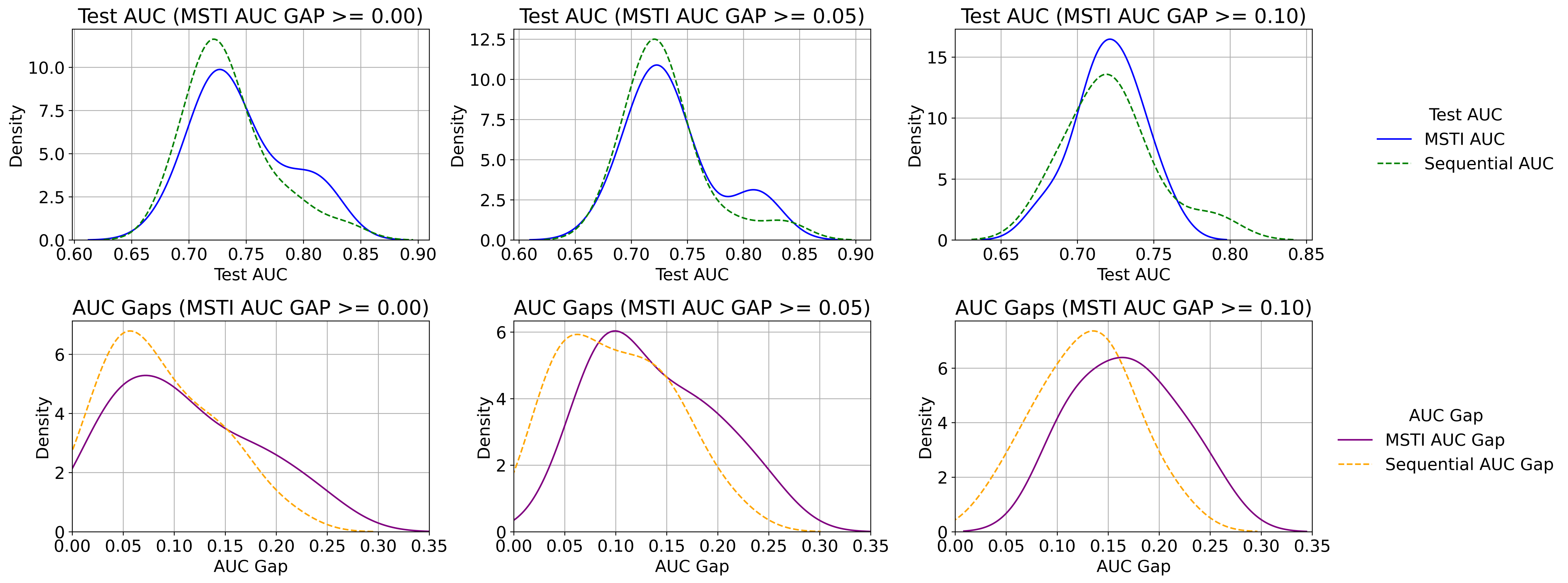}
    \caption{\textbf{Distribution of performance and fairness for MSTI model and sequential model.
    (a) Top: Test AUC distributions when the source is MSTI model and the sequential model under varying MSTI AUC gap thresholds ($>=$0.00, 0.05, and 0.10). (b) Bottom: AUC Gap distributions for MSTI model and sequential model under the same thresholds. Histograms are normalized to have an area of 1 (density).}}
    \label{fig:RQ2.2.1}
\end{figure*}

We compare the performance and fairness metrics of the sequential model with the MSTI model across three fairness thresholds: when the MSTI AUC gap exceeds 0.00, 0.05, and 0.10. Figure~\ref{fig:RQ2.2.1} shows that sequential training has the potential to improve fairness without compromising performance. The WTNDD test for differences in the Test AUC distributions passed under all three thresholds, indicating no significant differences in Test AUC distributions between the MSTI model and the sequential model. In contrast, the WTNDD test for differences in AUC Gap distributions failed, with Wasserstein Distances of 0.26, 0.37, and 0.46 for MSTI AUC Gaps exceeding 0.00, 0.05, and 0.10, respectively. This increasing Wasserstein Distance for the AUC Gap reflects a growing distributional difference, which indicates that the less fair the original model is (i.e., the larger the AUC Gap), the stronger the ability of sequential training to mitigate fairness issues. 
\begin{figure*}[h] %
    \centering
    \includegraphics[width=0.7\textwidth]{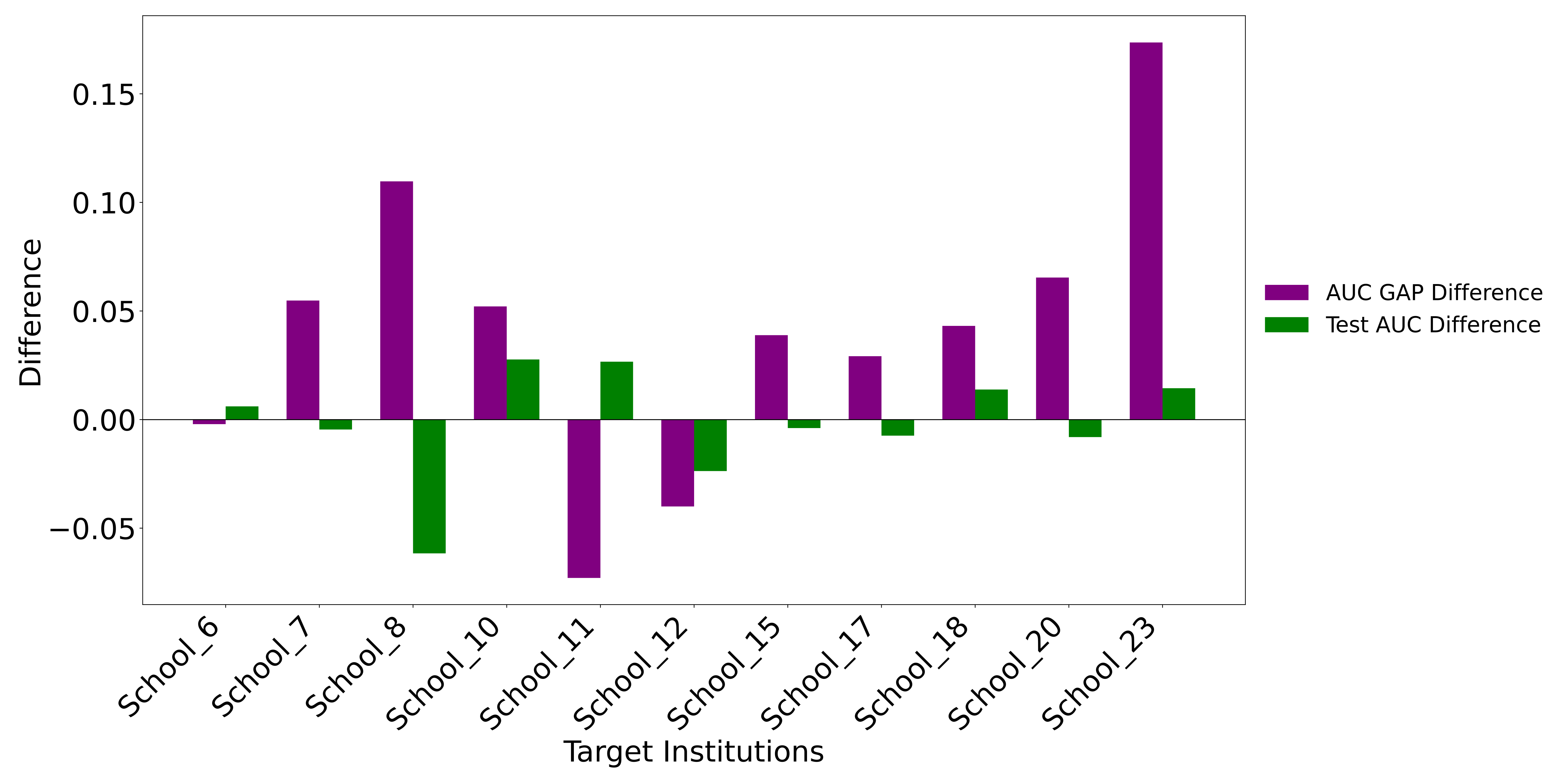}
    \caption{\textbf{Differences in AUC Gap and Test AUC between the MSTI model and the sequential model for each target institution when MSTI AUC Gap $>$ 0.10. A positive value indicates an improvement of tested metric with sequential training.}}
    \label{fig:RQ2.2.2}
\end{figure*}

From the perspective of individual target institutions, Figure~\ref{fig:RQ2.2.2} shows that for the majority of the selected target institutions (8 out of 11), the AUC Gap decreased when the sequential model was implemented compared to the MSTI model. However, this reduction in AUC Gap is occasionally accompanied by a performance degradation for some institutions. Overall, the results suggest that sequential training has the potential to improve model fairness without compromising performance, particularly for models with higher levels of initial unfairness. Nonetheless, there remains variance across individual target institutions, which underscores the need for thorough consideration of the specific contextual and institutional differences when applying sequential training in practice.

\subsection{Transfer Learning Strategies in Model Evaluation}
For our third research question, we aim to investigate how a target institution can mitigate the risks associated with deploying pre-trained models. As mentioned in Section 4.3, SFDA offers a strategy to adapt a model to a new domain without requiring access to either the original training data or the labeled target dataset.
\begin{figure*}[h] %
    \centering
    \includegraphics[width=0.9\textwidth]{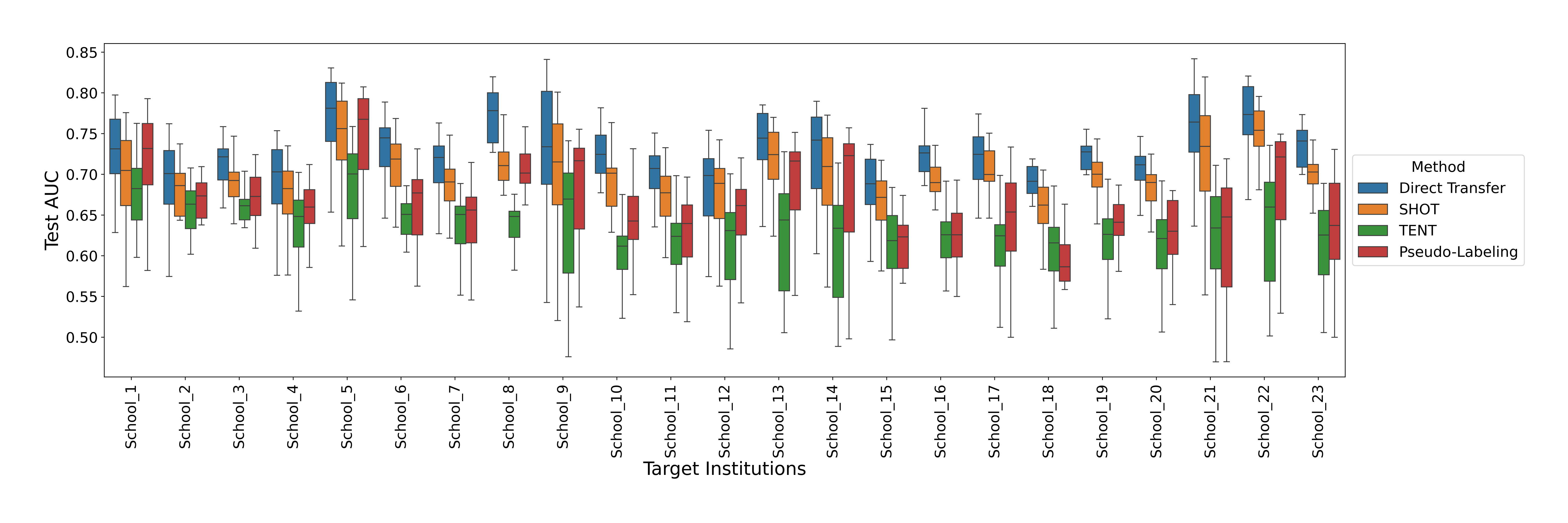}
    \caption{\textbf{Test AUC distributions across four different methods—Direct Transfer (blue), SHOT (orange), TENT (green), and Pseudo-Labeling (red)—for each target institution. Each box represents the interquartile range (IQR) of the AUC values, with the whiskers extending to the minimum and maximum values within 1.5 times the IQR.}}
    \label{fig:RQ3.1.1}
\end{figure*}
Surprisingly, Figure~\ref{fig:RQ3.1.1} shows that SFDA techniques that have demonstrated considerable success in computer vision tasks were not as effective in predicting student retention rates. In fact, many of these adaptation methods underperformed compared to directly applying the pre-trained models without any adaptation.

During the evaluation phase of pre-trained models for a target institution, we compare the use of two customized evaluation threshold methods discussed in Section 4.3 with the default threshold to assess their effectiveness.
We first compare the distributional differences in specificity and MCC of direct transfer models using the default threshold and the two optimal thresholds. 
\begin{figure*}[h]
    \centering
    \includegraphics[width=0.8\textwidth]{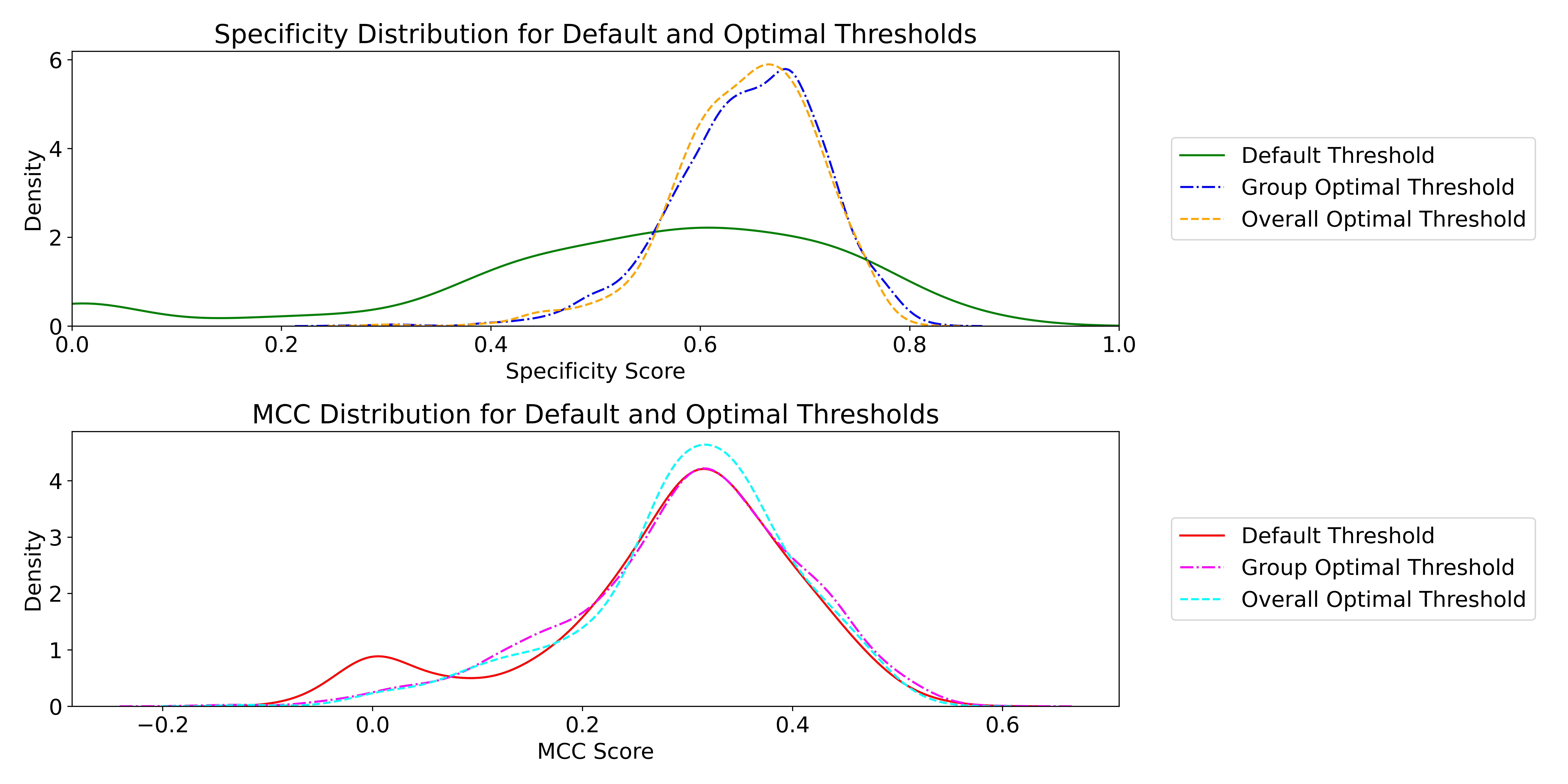}
    \caption{\textbf{Distribution of specificity (top) and MCC (bottom) scores of direct transfer models under the usage of default threshold, group-optimal threshold, and overall-optimal threshold.  Histograms are normalized to have an area of 1 (density). }}
    \label{fig:RQ3.2.2}
\end{figure*}
Figure~\ref{fig:RQ3.2.2} demonstrates that, compared to using the default threshold, both the group-optimal and overall-optimal threshold methods improve specificity, enhancing the model's ability to accurately identify dropout cases (WTNDD failed). Additionally, the WTNDD test passed when comparing the specificity distributions of group-optimal and overall-optimal thresholds, which indicates no substantial difference between them. In addition, there is no notable difference in the MCC distributions among using the group-optimal, overall-optimal, and the default thresholds (WTNDD passed).
\begin{figure*}[h] %
    \centering
    \includegraphics[width=0.8\textwidth]{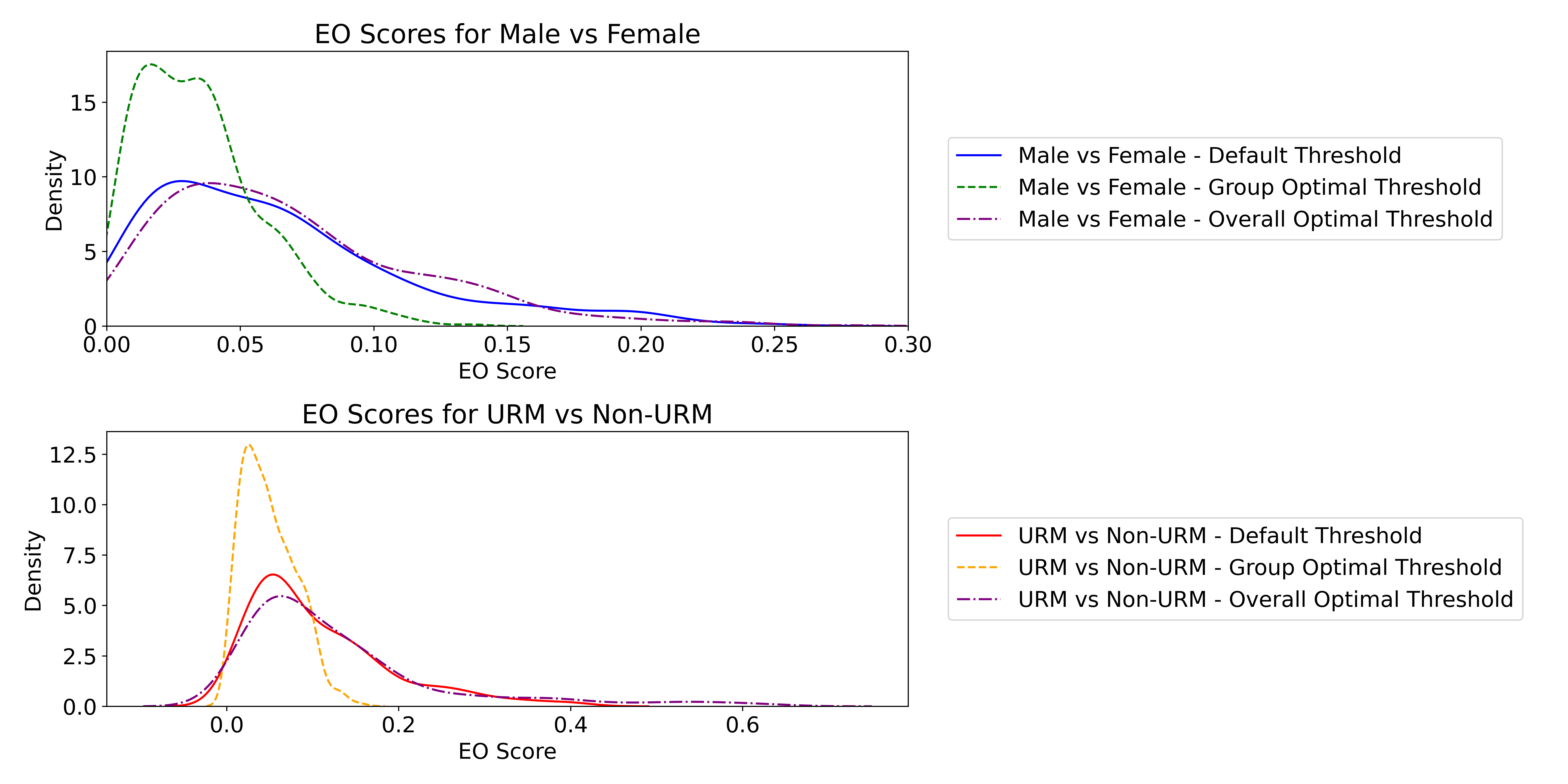}
    \caption{\textbf{Distribution of Equalized Odds (EO) scores of direct transfer models. (a) Top: EO scores for male vs. female under the usage of  default threshold, group optimal threshold, and overall optimal threshold. (b) Bottom: EO scores for URM vs. Non-URM under the usage of same thresholds.
    Lower score in EO suggests better model fairness. Histograms are normalized to have an area of 1 (density).}}
    \label{fig:RQ3.2.1}
\end{figure*}
In terms of fairness, Figure~\ref{fig:RQ3.2.1} presents the distributions of Equalized Odds (EO) scores of direct transfer models for different thresholds applied to male vs. female and URM vs. non-URM groups. The results indicate that the use of group-optimal thresholds enhances fairness, as evidenced by lower EO scores across gender and URM/non-URM groups compared to both the default and overall-optimal thresholds (WTNDD failed). There is no notable difference in the distribution of EO scores between using the default and overall optimal thresholds (WTNDD passed). Our results show that employing overall-optimal and group-optimal thresholds can effectively improve the model’s specificity without compromising overall performance, and employing group-optimal thresholds can further enhance fairness compared to overall-optimal thresholds.

\section{Discussion and Conclusion}

In this study, we examine the practical challenges and potential solutions for applying pre-trained models across institutions. We conduct the first large-scale empirical investigation of cross-institutional transfer learning of prediction models across both 4-year universities and 2-year community colleges. This study advances transfer learning in education by proposing actionable strategies to tackle transferability challenges identified in prior work \cite{xu2024contexts,li2021limitsalgorithmicpredictionglobe}. For example, our findings demonstrate that publicly available contextual information can effectively guide model selection and decision-making, even under privacy constraints.  Additionally, we mitigate fairness concerns by proposing potential solutions such as sequential training to improve fairness from a developer’s perspective and customizing evaluation thresholds for different groups to enhance fairness from a user’s perspective. These strategies align with broader efforts to mitigate algorithmic bias and promote equity in predictive modeling \cite{bird2024algorithms, kizilcec2021algorithmicfairnesseducation}. 

These findings have important implications. For model developers, our findings show that utilizing public contextual information facilitates model selection and can lay the groundwork for optimizing model choice under privacy constraints. 
For practitioners, our results demonstrate the feasibility of institutional collaboration under privacy constraints, showing that even without sharing raw data, it is possible to construct improved models through sequential training that benefits diverse population. 
For educational researchers, our study addresses a critical gap by focusing on the impact of cross-institutional transfer models on community colleges, a relatively under-explored area in the field. In addition, by providing empirical evidence of how these models perform across institutions with varying resource levels, this research sets the stage for future efforts to enable resource-rich schools to support under-resourced institutions through model transfer and adaptation.

This study has several limitations. First, the models rely on common data features across institutions, which may exclude some important indicators due to varying data infrastructures. Second, the sample includes only research university and  community college in the United States, which may limit the generalizability of the findings to other institutional types. Future research should address feature space differences between source and target institutions, expand the sample to include a wider range of institutions, and explore applications in other sectors, such as finance and healthcare, to further validate transfer learning frameworks in privacy- and fairness-critical contexts.

\begin{acks}

This work was supported by funding from the Learning Engineering Virtual Institute through the Fairness Analysis and Transfer Learning Hub. We extend our gratitude to Rene Kizilcec, Christopher Brooks, and Catherine Finnegan for their invaluable data and technical support.
\end{acks}

\bibliographystyle{ACM-Reference-Format}
\bibliography{mainbib}

\end{document}